# CAN COMPLEX NETWORKS DESCRIBE THE URBAN AND RURAL TROPOSPHERIC $O_3$ DYNAMICS?

Carmona-Cabezas Rafael[1,*], Gómez-Gómez Javier[1], Ariza-Villaverde Ana B.[1], Gutiérrez de Ravé Eduardo[1], Jiménez-Hornero Francisco J.[1]

[1]Department of Graphic Engineering and Geomatic, University of Cordoba, Gregor Mendel Building (3rd floor), Campus Rabanales, 14071 Cordoba, Spain

* Corresponding author. e-mail: f12carcr@uco.es

ABSTRACT

Tropospheric ozone ($O_3$) time series have been converted into complex networks through the recent so-called visibility graph (VG), using the data from air quality stations located in the western part of Andalusia (Spain). The aim is to differentiate the behavior between rural and urban regions when it comes to the ozone dynamics. To do so, some centrality parameters of the resulting complex networks have been investigated: the degree, betweenness and shortest path.

Results from these parameters coincide when describing the difference that tropospheric ozone exhibits seasonally and geographically. It is seen that ozone behavior is multifractal, in accordance to previous works. Also, it has been demonstrated that this methodology is able to characterize the divergence encountered between measurements in urban environments and countryside.

Additionally, the promising outcomes of this technique support the use of complex networks for the study of air pollutants dynamics, adding nuances to those reported by descriptive statistics or multifractal analysis.

1. INTRODUCTION

In the last few decades, tropospheric ozone has been the focus of many studies performed in different areas and scales around the world. This interest on ozone dynamics analysis and characterization has been awaken because it is one of the main photochemical oxidants on account of its abundance. When this irritant gas is found in high concentrations, severe impacts affect human health and harvest (Doherty et al., 2009). Those mentioned harms have a severe impact for economy, leading to losses of several billions of dollars annually (Miao et al., 2017).

The gas object of the presented work is a secondary pollutant and it is known that its creation and destruction mechanisms are related to photochemical and nonlinear processes (Graedel and Crutzen, 1993; Trainer et al., 2000). The mentioned processes depend on meteorological features such as wind direction, temperature, and principally solar radiation (Graedel and Crutzen, 1993; Guicherit and van Dop, 1977). They have been analyzed for the Mediterranean basin by some authors previously (Cieslik and Labatut, 1997; Güsten et al., 1994; Kouvarakis et al., 2000; Ribas and Peñuelas, 2004). Besides, tropospheric ozone concentration levels depends also on the presence of other gases, such as nitrogen oxides and volatile organic compounds (called precursors) that are produced in the urban and industrial areas (Sillman, 1999). Because of all these considerations the analysis of the temporal dynamics of ozone becomes a very complex task. As a consequence, traditional statistical analysis of ozone may offer a limited view of more complex dynamics of signals where the level of variability is high (Pavón-Domínguez et al., 2015).

One of the focuses on this topic is the difference encountered between rural and urban areas (García-Gómez et al., 2016; Jirik et al., 2017; Kumar et al., 2015). Despite the fact that ozone is created mainly in urban nuclei, it is proven that higher concentrations of this contaminant are measured within rural regions, leading to reduction of crop yields among other environmental problems (Tai and Val Martin, 2017). One reason attributed to this phenomenon is the transport of ozone by the wind to the less industrialized regions. In those areas the destruction rate of ozone is less significative than in urban ones.

In the last few years, a new technique to analyze temporal series has been developed (Lacasa et al., 2008). This technique relies in the use of complex networks obtained from the transformation of those signals (ground-level ozone concentration time series in this case). It was given the name of *Visibility Graph* (VG). These complex networks have proven to have several advantages such as: i) they inherit the features of the associated time series, which ends up resulting on additional feedback through the degree distribution (that will be defined later). ii) In addition, they can be used for analyzing series of several variables simultaneously, which could be very helpful for finding correlations between tropospheric ozone and its precursors; iii) and lastly, this recent bridge between complex networks and temporal series opens a wide range of new opportunities within the study of complex signals.

In the presented work, the complex networks obtained from ozone time series through the VG are used to retrieve the centrality parameters. These parameters give information about the most important nodes in a system and will be further explained in subsection 2.3. Finally, the main purpose is to check the ability of this methodology to analyze the differences in the behavior of the tropospheric ozone between urban and rural environments.

2. <u>MATERIALS AND METHODS</u>

   2.1. <u>Data</u>

The region that is the object of this analysis corresponds to the Guadalquivir Valley (South-western part of Spain) since the area has the proper orography, weather and anthropic conditions to be vulnerable to pollution by tropospheric ozone (Domínguez-López et al., 2014).

The data used here correspond to 1-hour monthly ozone concentration values collected from 2013 to 2017. The measurements were performed at four different stations located in the province of Cadiz (see Figure 1). Two of them are located in the southeastern part of the region (Algeciras and Alcornocales) and the others in the northwestern one (Cadiz and Prado del Rey). Algeciras and Cadiz correspond to urban areas, whereas the other two are situated within the natural reservoir named "Parque natural de Los Alcornocales", and so they have been labelled as rural.

These stations are part of the network that monitors the air pollution levels in the region of Andalusia, co-funded by the European Union and the Consejería de Medio Ambiente

y Ordenación del Territorio (Regional Environmental and Territory Management Department). The data was collected and provided lastly by the EEA (European Environmental Agency).

Figure 2 shows ozone concentration time series for four months from 2015 for two different locations: Alcornocales (rural) and Algeciras (urban). One clear difference between the two of them is that in the urban station, the lowest values measured are in many occasions close to zero. On the other hand, in the rural one that is not indeed the case: the ozone concentration does not vanish in the whole month or barely does it. This behaviour can be extended to all the rest of the years included in the presented study.

As it was clearly seen in Figure 2, the ozone concentrations reach especially high values in summer, where the case in winter is the exact opposite. Those differences were observed for this area (Adame et al., 2008; Jiménez-Hornero et al., 2010), and the reason is that the most suitable conditions for the ozone creation are found around summer. The temperature and solar radiation progressively raise and reach their peak in July, which allows higher creation rates and therefore concentration. One of the reactions that governs this mechanism is the following (Graedel and Crutzen, 1993).

$$NO_2 + O_2 \leftrightarrow O_3 + NO \qquad (1)$$

This photochemical reaction is reversible and tends to the ozone creation (rightwards direction) when there is energy (light) available and vice versa when there is not. For that reason, the highest and lowest values of ozone are always measured during the day and night, respectively. The same happens summer and winter, as discussed previously.

2.2. Visibility graphs

One possible definition for a graph is a set of points vertices or nodes that are connected through lines called *edges*. As commented above, a tool that makes possible the transformation of a time series into a graph was introduced by Lacasa et al. (2008) and called Visibility Graph (VG). One of the main features that rises the interest of researchers relies on the fact that it inherits many properties of the original series.

The first thing that must be done for constructing the visibility adjacency matrix (which contains all the information of the new network), is to stablish a method to decide which points (or nodes) in the system are connected to each other or have visibility. The criterion is the following: two arbitrary points from the time series ($t_a$, $y_a$) and ($t_b$, $y_b$) will have visibility (and will be connected in the graph) if any given point ($t_c$, $y_c$) that is located in between ($t_a < t_c < t_b$) fulfills the following condition:

$$y_c < y_a + (y_b - y_a)\frac{t_c - t_a}{t_b - t_a} \qquad (2)$$

One example of how this method works can be seen in Figure 3, where it is applied to a sample time series as an illustration. It can be observed that the original temporal series has been transformed into a complex network, where some of the points are connected by edges. This new graph will inherit the complexity of the original series (Lacasa et al., 2008; Lacasa and Toral, 2010), meaning that a regular graph would be created from a periodic time series, for instance.

After applying this visibility method, the result is a NxN adjacency binary matrix, with N the total number of points in the system. The information of the nodes is given by each row of the matrix, so that $a_{ij} = 1$ means that the node $i$ and $j$ have visibility; whereas $a_{ij} = 0$ means the opposite case (no edge connects those two nodes). The algorithm can be substantially simplified (reducing the computational cost of the process) if some considerations are considered. These can be done thanks to the properties that the adjacency matrix holds. The properties are listed below:

- Hollow matrix: Since there are no intermediate nodes to fulfill the condition, in the case of the diagonal, all the elements are zero ($a_{ii} = 0$). Hence a node does not have visibility with itself.

- Symmetric matrix: Due to the reciprocity of the visibility between two nodes, all the nodes in system fulfill $a_{ij} = a_{ji}$. This is a property of all undirected graphs.

- Nearest neighbors: The elements that surround the diagonal are always 1 ($a_{ij} = 1$ for $j = i \pm 1$). This is because each point always sees the closest previous and next node (there are no points in between to prevent the visibility).

With all these considerations, every visibility adjacency matrix has a general form as follows:

$$A = \begin{pmatrix} 0 & 1 & \cdots & a_{1,N} \\ 1 & 0 & 1 & \vdots \\ \vdots & 1 & \ddots & 1 \\ a_{N,1} & \cdots & 1 & 0 \end{pmatrix} \quad (3)$$

2.3. Centrality measures

When trying to retrieve information from a given complex network, one of the most commonly used approaches is discerning which of them are the most important nodes in the system. To this purpose, centrality measures comes usually in handy. This concept was initially applied to the study of social networks and later transferred to other fields of knowledge (Agryzkov et al., 2019; Joyce et al., 2010; Liu et al., 2015). This work has been focused on two of them: the degree and betweenness centrality measures, which will be explained afterwards.

### 2.3.1. Degree centrality

A possible definition for the degree of a node ($k_i$) is the number of other nodes that have visibility with it ($k_i = \sum_j a_{ij}$). For instance, in Figure 3, the degree of the three first nodes are $k = 3$, $k = 2$ and $k = 3$, respectively. On the whole, it is possible to obtain the probability that corresponds to each degree, by simply counting how many times each value is repeated. From there, the degree distribution of the sample $P(k)$ can be retrieved.

By analyzing the degree distribution that is built from the VG it is possible to describe the nature of the time series, as previous works have shown (Lacasa et al., 2008; Mali et al., 2018). It has been probed its capability to distinguish between fractal, random or periodic signals, for instance. Thus, by studying the degree distribution, a first insight of the behaviour of the ozone concentration time series can be yielded as first step before getting into a more complex analysis. As some previous works explain (Lacasa et al., 2009; Lacasa and Toral, 2010), time series which have VGs whose degree distributions can be fitted to a power law $P(k) \propto k^{-\gamma}$ correspond to scale free due to the effect of *hub* repulsion (Song et al., 2006). The term *hub* refers to the nodes with unlikely highest number of links (highest degrees, see Figure 4). The right tail of each degree distribution, governed by those *hubs*, can be represented in a log-log plot and fitted by a simple linear regression. The slope obtained by this regression provides an interesting parameter, the so-called $\gamma$ exponent, which has already been used in some works (Lacasa and Toral, 2010; Mali et al., 2018). In Figure 4a, thanks to the v-k plots (Pierini et al., 2012), it is possible to appreciate how *hubs* from the VG are related to the largest values of ozone concentration.

### 2.3.2. Betweenness centrality

Before presenting this quantity, it is necessary to introduce a definition for the shortest path (SP). It can be understood easily that the SP for a pair of nodes ($i$, $j$) in a VG is the minimum number of edges between both. Consequently, the SP between two consecutive nodes will be the unit.

The betweenness of a node ($b_i$) for an undirected graph is defined as the total number of SPs which passes through this node and mathematically:

$$b_i = \sum_{\substack{j=1 \\ j \neq i}}^{N} \sum_{\substack{k=1 \\ k \neq i,j}}^{N} \frac{n_{jk}(i)}{2N_{jk}} \quad (4)$$

where $N_{jk}$ is the total number of SPs from node $j$ to node $k$ and $n_{jk}(i)$ is the number of SPs from node $j$ to $k$, that contains the node $i$. It is divided by two in order not to repeat the same pair of nodes twice in an undirected graph (the path from $j$ to $k$ and vice versa give the same information).

The mentioned parameter estimates the centrality of a node by considering whether it is between many of the nodes (Latora et al., 2017). In an equivalent way to the degree *hubs*, some points with a remarkably higher betweenness exist. For the seek of clarity, authors propose the term *skyline hubs* to refer to those with unlikely high betweenness. This name has been chosen because of its similarity to the skyline drawn by the skyscrapers in a city. Those temporal nodes are characterized by being the points which the SPs of many other nodes will pass necessarily through. In Figure 4b it is possible to see how they are related to *hubs* and therefore to some peaks of the time series. They are a more selective way to identify key nodes in the signal.

3. RESULTS AND DISCUSSION

The first step was to transform the ozone concentration time series from all the locations for all the months (from 2013 to 2017) into complex networks with the VG algorithm. In this section, the results that are shown and discussed were obtained from the direct analysis of these networks.

3.1. Degree centrality

Following the definition of degree ($k$) given previously (subsection 2.3.1), the number of edges connected to each node in the different VGs was computed. From these values, it has been possible to construct the degree distribution of the networks, as shown in Figure 5. In this plot, the degree distribution of the whole year (2015 shown as reference) was computed; with the final aim of discerning whether this distribution could give some insight on the difference between ozone dynamics in rural and urbanized areas. This first guess was motivated by previous works that use it in order to analyze the nature of the time series for several quantities (Lacasa and Toral, 2010; Mali et al., 2018).

The results for all the years and months are quite similar indeed, in accordance with previous studies (Carmona-Cabezas et al., 2019); and for that reason, only one year (2015) is used for the sake of clarity in Figure 5. As can be seen in the cited plot, the tail of the distributions follows a power law of the form $P(k) \propto k^{-\gamma}$, that leads to a linear part of the curve when plotted in logarithmic scale both $k$ and $P(k)$. This behaviour points to the fractal nature of the signal, which was expected looking at some prior analyses (Jiménez-Hornero et al., 2010; Pavon-Dominguez et al., 2013). The slope of the linear portion in absolute value leads to the computation of this $\gamma$ parameter. It is clear that the trend is negative, since the nodes with biggest degrees and known as *hubs* correspond usually to the high values of the distribution (see Figure 4), whose likeliness is very low. In all the cases studied here, the distributions are almost overlapped and finally the exponent $\gamma \sim 3.4$ roughly for all of them (see Table 1), as depicted in Figure 5. For that reason, this parameter alone is not able to distinguish the dynamics of the tropospheric ozone in the different regions on which this study focuses (urban - rural). Nevertheless, it does give useful information about the nature of the time series as discussed and validates the data, since equivalent studies for different years and geographical area gave similar values of $\gamma$ (Carmona-Cabezas et al., 2019).

Looking at the average degree ($\bar{k}$) of all the nodes from the VGs of each month, some information can be drawn. In Figure 6a), this averaged value is plotted for each month and the first thing that can be commented is that the shape of the curves changes along the year. For all the studied locations, there is an increasing tendency towards summer that then decays typically after August. This behaviour was expected as $\bar{k}$ would mean a higher number of *hubs* in the signal and those are related to the greatest concentrations of ozone as shown in Figure 4. That is in the end due to the more suitable conditions for ozone formation that exist in summer with respect to the other season (specially winter). One interesting thing that was as well observed by the authors in a recent study (Carmona-Cabezas et al., 2019) is the fact that this quantity drops around April and November. One possible explanation is the weather of spring and autumn, unstable by nature, that favors the dispersion of gases and particles in the air (tropospheric ozone amongst them), as discussed in other studies (Dueñas et al., 2002).

Also, looking at Figure 6a), a clear difference between the curves of the rural (blue) and urban (red) areas is found. Having all the same behaviour mentioned before, the values of Prado del Rey and Alcornocales (both rural) are sensibly higher than those of Algeciras and Cadiz (urban). The difference between summer and winter is as well more pronounced in the rural locations. Authors attribute this fact to the transport of ozone with the wind (Dueñas et al., 2004), added to the process of destruction of that secondary contaminant, rather than its formation. This effect would correspond to the leftwards direction of the photochemical reaction described before (Equation 1). After the ozone is created, it starts to react with the Nitrogen Oxide during night conditions (absence of light and lower temperatures). As could be observed in Figure 2 and Figure 4, the ozone values reach minima of zero quite often in the urban area of Algeciras for instance, which is not the case for the rural location of Alcornocales. That is directly related to the higher concentrations of NO from factories and vehicle emissions that can be found in an industrial area such as Algeciras; in contraposition to the natural reservoir of Parque natural de los Alcornocales. Therefore, the ozone that is created and transported to the rural areas studied here cannot be transformed to other gases at the same rate as in the city, leading to higher concentrations values on average. This can be observed as well in Table *1*.

Another parameter from the computed degrees that can be seen (Figure 6b) is the standard deviation of $k$ ($\sigma_k$). As it was discussed in a previous work on this topic (Carmona-Cabezas et al., 2019), it is related with the differences encountered between daily and night values of ozone concentration. As expected, maxima are found again for summer, seemingly due to the fact that the biggest differences in UV radiation and temperature in this area are found in that season between day and night. Again, drops of this quantity $\sigma_k$ are observed in both spring and autumn, which authors attribute to the same reason as in the case of $\bar{k}$.

Now the curves of $\sigma_k$ in rural and urban areas are more similar than in the case $\bar{k}$, but still a difference can be observed. This lower difference with respect to the curves of $\bar{k}$ can be explained using the same effect discussed before: the destruction of ozone is lower in the rural areas.

### 3.2. Betweenness centrality

After the mentioned analysis of the degree, authors checked the suitability of the next centrality parameter: betweenness ($b$). Moreover, the average SP quantity was obtained in order to get more information about the time series, since it is directly related to $b$ (see Equation 4). Both of them were computed for every node in the VGs (from each month). After the information of each node was retrieved, the average was taken for SP, whereas in the case of the betweenness centrality, the median has been chosen. This decision was motivated by the fact that the distribution of the betweenness is much more skewed than those of the degree centrality and SP. In addition to that, the vast majority of values of $b$ are zero or very close to it (see Figure 4). For those reasons, authors consider median as a more representative measure of the overall behaviour rather than the mean. Results for $b$ and SP can be observed in Figure 7, where the different locations studied can be seen.

On the one hand, in Figure 7a) the betweenness centrality shows a seasonal pattern as well as the degree, being this one more pronounced in the rural areas than in the urban ones. As can be easily seen, the minimum values are reached for late spring, summer and early autumn (from May to October), in contrast to the degree centrality, which was maximum for this period. The reason of these minima is that the higher are the concentrations of ozone, the greater will be the amount of degree *hubs* and *skyline hubs*, as it was shown in Figure 4a). Since *skyline hubs* allow faster connections between nodes, less edges will be necessary to link them through the SP. And so, the average of this amount will be reduced over this period and vice versa for the rest of the year. This reduction in the average SP is seen in Figure 7b). By definition, the shorter is the SP between two points, the less nodes will be necessary for them to pass through, resulting on a lower betweenness in general (what is indeed happening in summer).

When it comes to the differences encountered in $b$ between rural and urban environments, which is not shown in the SP, authors attribute this effect to the degeneracy of the SP. In the computation of the average of this quantity, degeneracy is not taken into account because only the length is used and not the number of possible SP between two given nodes. That is not the case for the betweenness, whose definition is based on this degeneracy $N_{jk}$ (see Equation 4). The larger $N_{jk}$, the lower will be the resulting value of betweenness for each node, leading to a final lower median for all months and vice versa. Authors consider that the difference in the degeneracy among areas can be related to the dynamics of tropospheric ozone for diurnal values. That is because SPs always use mainly the highest values (*skyline hubs*) to cover most of the distance between two nodes. As a result, a higher degeneracy can be interpreted as a signal with a smoother envelope, because more options will be available to construct SP with same lengths. The opposite for the case of irregular envelope can be argued using the same reasoning.

4. <u>CONCLUSIONS</u>

On the whole, results show that the use of complex networks for analyzing temporal series of tropospheric ozone is suitable to distinguish the dynamics in rural and urban areas. The probability distribution of the degree centrality $P(k)$ identifies the nature of the signals, being fractal for all the cases, as it was previously known (Jiménez-Hornero et al., 2010; Pavon-Dominguez et al., 2013). Moreover, by looking at the values of $\bar{k}$ and $\sigma_k$ a seasonal behaviour has been observed. Besides, clear differences between rural and urban locations can be appreciated from those values, specially in the case of the average degree. Betweenness centrality has turned out to be a supplementary source of information for diurnal behaviour (envelope of the signals) and differences among the studied locations. All these outcomes support the capability of complex network analysis to describe ozone dynamics and transport from the urban to the rural environments.

To conclude, the advantages of using VGs for the analysis of time series and particularly from tropospheric ozone must be emphasized. In the last years, advances in the field of complex networks have made them a very convenient tool for several reasons: their computation efficiency, suitability for big data series and wide range of application, among others. In addition to that, since VG is a state-of-the-art methodology, it opens multiple possibilities for future works. Authors consider appropriate to focus on the use of *multiplex visibility graphs* (Lacasa et al., 2015) to study multivariate time series. Furthermore, the concept of *skyline hubs* could be employed to identify relevant points in a time series, leading to different ways of understanding the betweenness centrality parameter applied to time series, similarly to degree *hubs*.

5. ACKNOWLEDGEMENTS

The FLAE approach for the sequence of authors is applied in this work. Authors gratefully acknowledge the support of the Andalusian Research Plan Group TEP-957 and the XXIII research program (2018) of the University of Cordoba. R. Carmona-Cabezas truly thanks the backing of the "Programa de Empleo Joven" (European Regional Development Fund / Andalusia Regional Government).

6. REFERENCES

Adame, J.A., Lozano, A., Bolívar, J.P., De la Morena, B.A., Contreras, J., Godoy, F., 2008. Behavior, distribution and variability of surface ozone at an arid region in the south of Iberian Peninsula (Seville, Spain). Chemosphere 70, 841–849. https://doi.org/10.1016/j.chemosphere.2007.07.009

Agryzkov, T., Tortosa, L., Vicent, J.F., 2019. A variant of the current flow betweenness centrality and its application in urban networks. Appl. Math. Comput. 347, 600–615. https://doi.org/10.1016/j.amc.2018.11.032

Carmona-Cabezas, R., Ariza-Villaverde, A.B., Gutiérrez de Ravé, E., Jiménez-Hornero, F.J., 2019. Visibility graphs of ground-level ozone time series: A multifractal analysis. Sci. Total Environ. 661, 138–147. https://doi.org/10.1016/j.scitotenv.2019.01.147

Cieslik, S., Labatut, A., 1997. Ozone and heat fluxes over a Mediterranean pseudosteppe. Atmos. Environ. 31, 177–184. https://doi.org/10.1016/S1352-2310(97)00084-8

Doherty, R.M., Heal, M.R., Wilkinson, P., Pattenden, S., Vieno, M., Armstrong, B., Atkinson, R., Chalabi, Z., Kovats, S., Milojevic, A., Stevenson, D.S., 2009. Current and future climate- and air pollution-mediated impacts on human health. Environ. Health 8, S8. https://doi.org/10.1186/1476-069X-8-S1-S8

Domínguez-López, D., Adame, J.A., Hernández-Ceballos, M.A., Vaca, F., De la Morena, B.A., Bolívar, J.P., 2014. Spatial and temporal variation of surface ozone, NO and NO2 at urban, suburban, rural and industrial sites in the southwest of the Iberian Peninsula. Environ. Monit. Assess. 186, 5337–5351. https://doi.org/10.1007/s10661-014-3783-9

Dueñas, C., Fernández, M.., Cañete, S., Carretero, J., Liger, E., 2004. Analyses of ozone in urban and rural sites in Málaga (Spain). Chemosphere 56, 631–639. https://doi.org/10.1016/j.chemosphere.2004.04.013

Dueñas, C., Fernández, M.., Cañete, S., Carretero, J., Liger, E., 2002. Assessment of ozone variations and meteorological effects in an urban area in the Mediterranean Coast. Sci. Total Environ. 299, 97–113. https://doi.org/10.1016/S0048-9697(02)00251-6

García-Gómez, H., Aguillaume, L., Izquieta-Rojano, S., Valiño, F., Àvila, A., Elustondo, D., Santamaría, J.M., Alastuey, A., Calvete-Sogo, H., González-Fernández, I., Alonso, R., 2016. Atmospheric pollutants in peri-urban forests of Quercus ilex: evidence of pollution abatement and threats for vegetation. Environ. Sci. Pollut. Res. 23, 6400–6413. https://doi.org/10.1007/s11356-015-5862-z

Graedel, T.E., Crutzen, P.J., 1993. Atmospheric change: an earth system perspective. J. Chem. Educ. https://doi.org/10.1021/ed070pA252.2

Guicherit, R., van Dop, H., 1977. Photochemical production of ozone in Western Europe (1971–1975) and its relation to meteorology. Atmos. Environ. 11, 145–155. https://doi.org/10.1016/0004-6981(77)90219-0

Güsten, H., Heinrich, G., Weppner, J., Abdel-Aal, M.M., Abdel-Hay, F.A., Ramadan, A.B., Tawfik, F.S., Ahmed, D.M., Hassan, G.K.Y., Cvitaš, T., Jeftić, J., Klasinc, L., 1994.

Ozone formation in the greater Cairo area. Sci. Total Environ. 155, 285–295. https://doi.org/10.1016/0048-9697(94)90507-X

Jiménez-Hornero, F.J., Gutiérrez de Ravé, E., Ariza-Villarverde, A.B., Giráldez, J.V., 2010. Description of the seasonal pattern in ozone concentration time series by using the strange attractor multifractal formalism. Environ. Monit. Assess. 160, 229–236. https://doi.org/10.1007/s10661-008-0690-y

Jirik, V., Brezna, B., Machaczka, O., Honkysova, S., Miturova, H., Janout, V., 2017. Associations between air pollution in the industrial and suburban parts of Ostrava city and their use. Environ. Monit. Assess. 189. https://doi.org/10.1007/s10661-017-6094-0

Joyce, K.E., Laurienti, P.J., Burdette, J.H., Hayasaka, S., 2010. A New Measure of Centrality for Brain Networks. PLoS ONE 5. https://doi.org/10.1371/journal.pone.0012200

Kouvarakis, G., Tsigaridis, K., Kanakidou, M., Mihalopoulos, N., 2000. Temporal variations of surface regional background ozone over Crete Island in the southeast Mediterranean. J. Geophys. Res. Atmospheres 105, 4399–4407. https://doi.org/10.1029/1999JD900984

Kumar, A., Singh, D., Singh, B.P., Singh, M., Anandam, K., Kumar, K., Jain, V.K., 2015. Spatial and temporal variability of surface ozone and nitrogen oxides in urban and rural ambient air of Delhi-NCR, India. Air Qual. Atmosphere Health 8, 391–399. https://doi.org/10.1007/s11869-014-0309-0

Lacasa, L., Luque, B., Ballesteros, F., Luque, J., Nuño, J.C., 2008. From time series to complex networks: The visibility graph. Proc. Natl. Acad. Sci. 105, 4972–4975. https://doi.org/10.1073/pnas.0709247105

Lacasa, L., Luque, B., Luque, J., Nuño, J.C., 2009. The visibility graph: A new method for estimating the Hurst exponent of fractional Brownian motion. Europhys. Lett. 86, 30001. https://doi.org/10.1209/0295-5075/86/30001

Lacasa, L., Nicosia, V., Latora, V., 2015. Network structure of multivariate time series. Sci. Rep. 5, 15508. https://doi.org/10.1038/srep15508

Lacasa, L., Toral, R., 2010. Description of stochastic and chaotic series using visibility graphs. Phys. Rev. E 82, 036120. https://doi.org/10.1103/PhysRevE.82.036120

Latora, V., Nicosia, V., Russo, G., 2017. Complex Networks [WWW Document]. Camb. Core. https://doi.org/10.1017/9781316216002

Liu, C., Zhan, X.-X., Zhang, Z.-K., Sun, G.-Q., Hui, P.M., 2015. Events Determine Spreading Patterns: Information Transmission via Internal and External Influences on Social Networks. New J. Phys. 17, 113045. https://doi.org/10.1088/1367-2630/17/11/113045

Mali, P., Manna, S.K., Mukhopadhyay, A., Haldar, P.K., Singh, G., 2018. Multifractal analysis of multiparticle emission data in the framework of visibility graph and sandbox algorithm. Phys. Stat. Mech. Its Appl. 493, 253–266. https://doi.org/10.1016/j.physa.2017.10.015

Miao, W., Huang, X., Song, Y., 2017. An economic assessment of the health effects and crop yield losses caused by air pollution in mainland China. J. Environ. Sci. 56, 102–113. https://doi.org/10.1016/j.jes.2016.08.024

Pavón-Domínguez, P., Jiménez-Hornero, F.J., Gutiérrez de Ravé, E., 2015. Joint multifractal analysis of the influence of temperature and nitrogen dioxide on tropospheric ozone. Stoch. Environ. Res. Risk Assess. 29, 1881–1889. https://doi.org/10.1007/s00477-014-0973-5

Pavon-Dominguez, P., Jimenez-Hornero, F.J., Gutierrez de Rave, E., 2013. Multifractal analysis of ground–level ozone concentrations at urban, suburban and rural background monitoring sites in Southwestern Iberian Peninsula. Atmospheric Pollut. Res. 4, 229–237. https://doi.org/10.5094/APR.2013.024

Pierini, J.O., Lovallo, M., Telesca, L., 2012. Visibility graph analysis of wind speed records measured in central Argentina. Phys. Stat. Mech. Its Appl. 391, 5041–5048. https://doi.org/10.1016/j.physa.2012.05.049

Ribas, À., Peñuelas, J., 2004. Temporal patterns of surface ozone levels in different habitats of the North Western Mediterranean basin. Atmos. Environ. 38, 985–992. https://doi.org/10.1016/j.atmosenv.2003.10.045

Sillman, S., 1999. The relation between ozone, NOx and hydrocarbons in urban and polluted rural environments. Atmos. Environ. 33, 1821–1845. https://doi.org/10.1016/S1352-2310(98)00345-8

Song, C., Havlin, S., Makse, H.A., 2006. Origins of fractality in the growth of complex networks. Nat. Phys. 2, 275–281. https://doi.org/10.1038/nphys266

Tai, A.P.K., Val Martin, M., 2017. Impacts of ozone air pollution and temperature extremes on crop yields: Spatial variability, adaptation and implications for future food security. Atmos. Environ. 169, 11–21. https://doi.org/10.1016/j.atmosenv.2017.09.002

Trainer, M., Parrish, D.D., Goldan, P.D., Roberts, J., Fehsenfeld, F.C., 2000. Review of observation-based analysis of the regional factors influencing ozone concentrations. Atmos. Environ. 34, 2045–2061.

World Health Organization, 2005. WHO Air quality guidelines for particulate matter, ozone, nitrogen dioxide and sulfur dioxide.

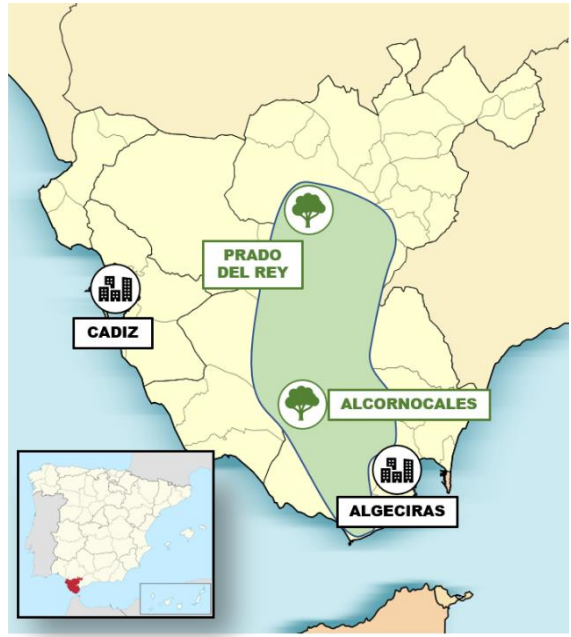

Figure 1: Location of the air quality control stations from which the data was retrieved. The image in the left-bottom corner shows the position of the studied area in the Iberian Peninsula. Green area indicates the natural reservoir "Parque natural de Los Alcornocales".

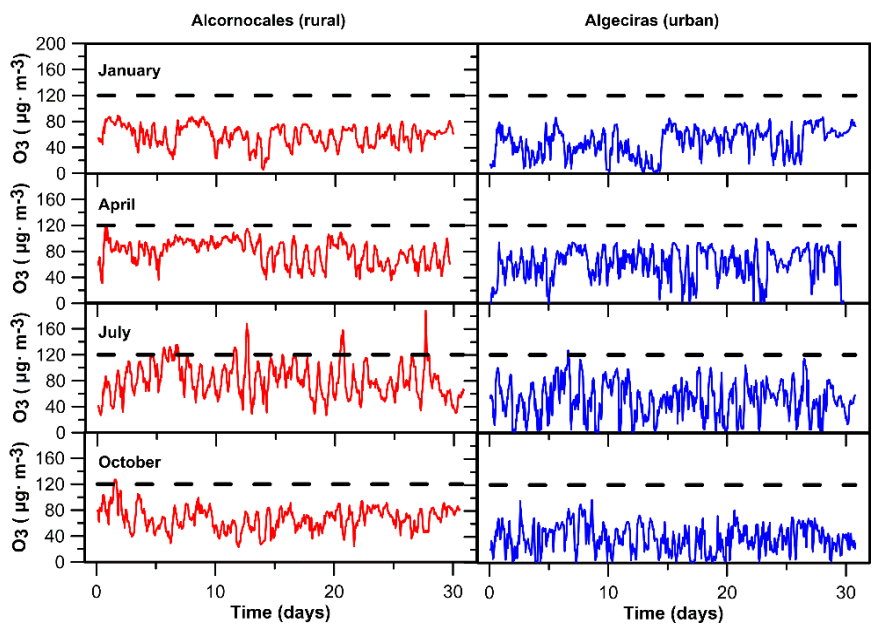

Figure 2: Examples of ozone concentration time series for four months and a given year (2015) in two locations, one rural (red) and the other urban (blue). The dashed line indicates the value 120 μg/m$^3$ stablished as a reference (World Health Organization, 2005).

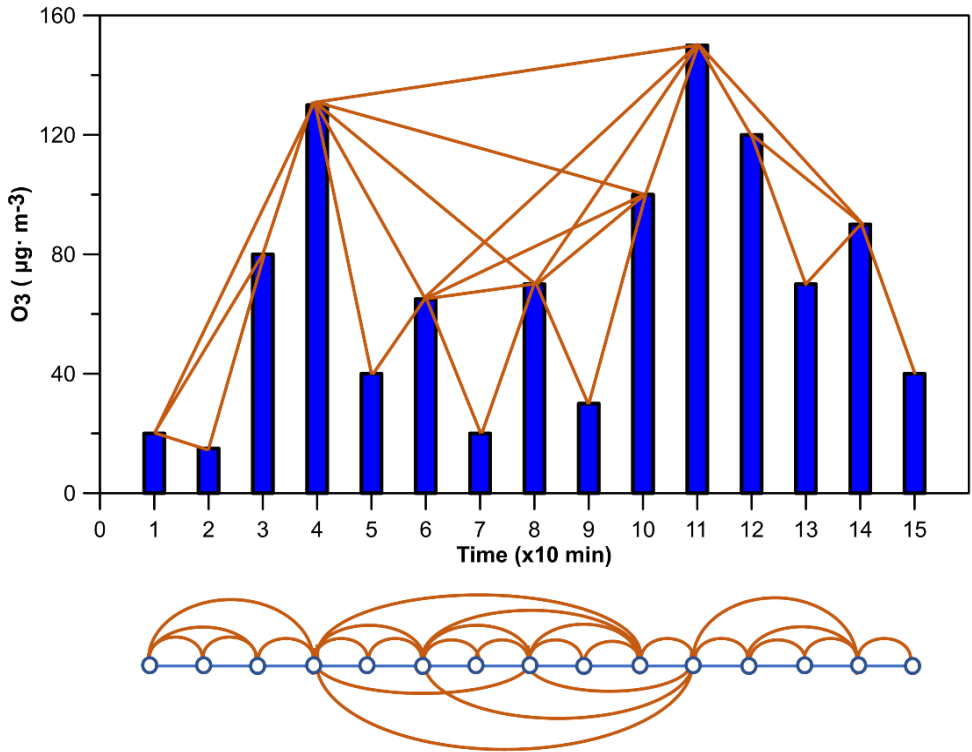

Figure 3: Sample time series transformed into a complex network through the visibility graph algorithm. Below, all the connections are shown in a more visual way.

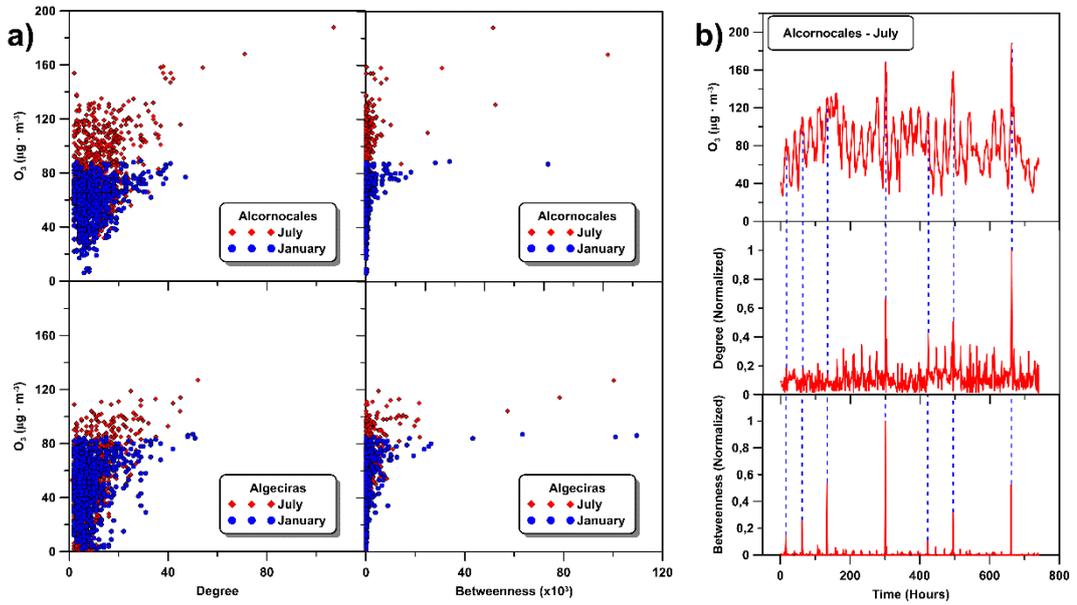

Figure 4: a) Plots of the values of tropospheric ozone concentration against the degree and betweenness of each point. b) Temporal distribution of these two quantities and the ozone concentration for a given month and location. Blue dashed lines in b) are used to associate several peaks.

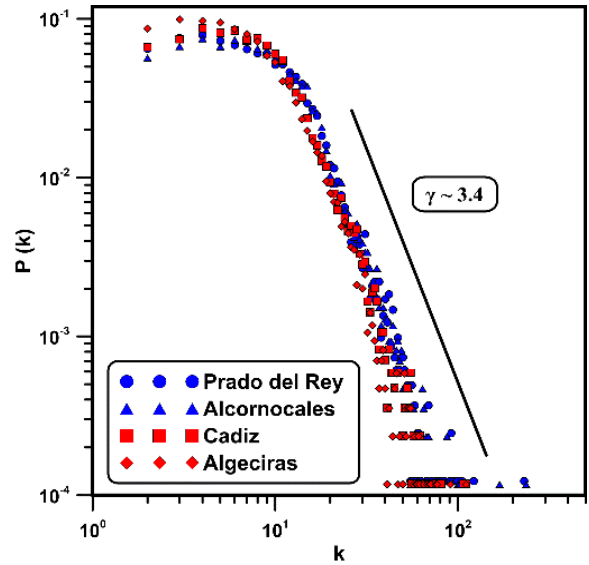

Figure 5: Annually degree distribution for the four places studied. The shown year corresponds to 2015 as an example, since all the other years have been checked to give equivalent information. The red and blue colors refer to urban and rural environments, respectively.

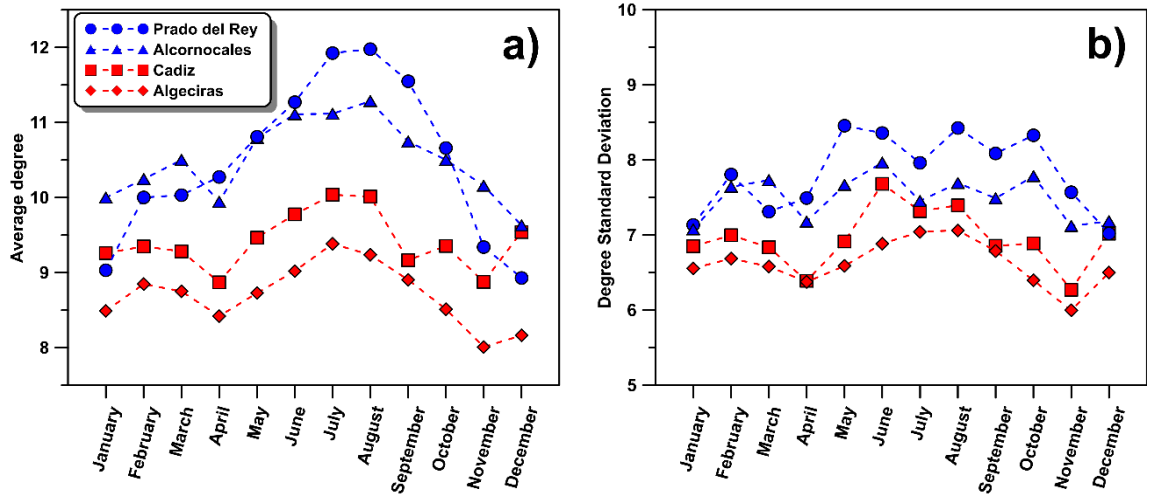

Figure 6: Computed average degree and standard deviation from the degree distribution of each month in the four locations considered. Each monthly value is the average of the computed ones from all the years available.

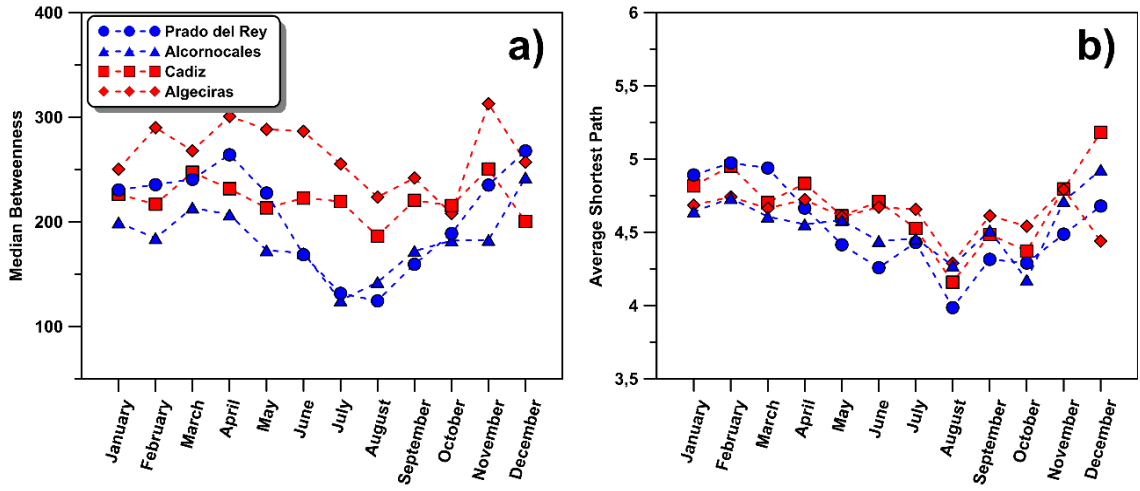

Figure 7: a) Median betweenness and b) average SP computed from the VG of each month in the four locations. Blue values indicate rural area, while the red ones are urban. Each monthly value is the average of the computed ones from all the years available.

| | Location | $\overline{O_3}(\mu g \cdot m^{-3})$ | $\bar{\gamma}$ |
|---|---|---|---|
| Northwestern coast | Prado del Rey | $81 \pm 4$ | $3.50 \pm 0.13$ |
| | Cadiz | $69 \pm 3$ | $3.37 \pm 0.25$ |
| Southeastern coast | Alcornocales | $72 \pm 3$ | $3.39 \pm 0.15$ |
| | Algeciras | $54 \pm 5$ | $3.40 \pm 0.12$ |

Table 1: Mean concentration and gamma exponent for each location (averaged for all the years).

**HIGHLIGHTS**

- Ozone time series are converted to complex networks through the visibility graph.
- Centrality measures are used to acquire information from the complex networks.
- *Skyline hubs* are introduced as a tool to identify relevant points in a signal.
- Urban-rural differences are exposed looking at degree and betweenness values.

**Graphical Abstract**

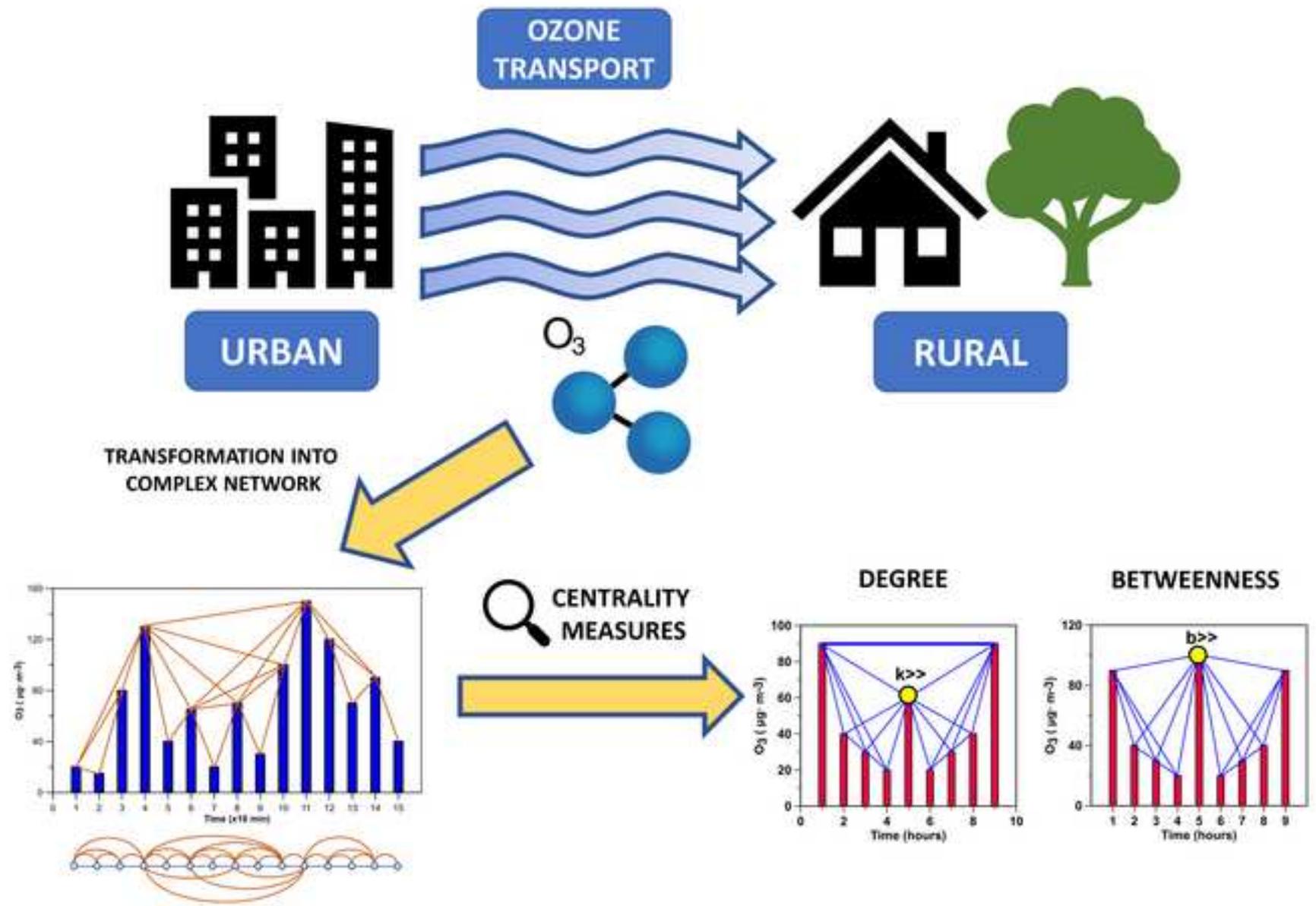